\documentclass[11pt]{elsart}
\usepackage{amsmath,amssymb}
\usepackage{epsfig}
\usepackage{algorithm}
\usepackage{algorithmic}
\usepackage{color}
\newcommand{\br}{\boldsymbol{r}}
\newcommand{\bc}{\boldsymbol{c}}
\newcommand{\bv}{\boldsymbol{v}}
\newcommand{\bV}{\boldsymbol{V}}
\newcommand{\bu}{\boldsymbol{u}}
\newcommand{\bkhat}{\boldsymbol{\hat k}}

\newcommand{\bB}{\boldsymbol{B}}
\newcommand{\bi}{\boldsymbol{i}}
\newcommand{\bj}{\boldsymbol{j}}

\newcommand{\cuda}{$\mbox{CUDA}^{\mbox{\tiny TM}}$ }
\newcommand{\nvidia}{$\mbox{NVIDIA}^\circledR$}

\begin{document}

\begin{frontmatter}

\title{Solving the Boltzmann Equantion on GPUs}

\author{A. Frezzotti},
\ead{aldo.frezzotti@polimi.it}
\author{G. P. Ghiroldi},
\ead{gian.ghiroldi@mail.polimi.it}
\author{L. Gibelli\corauthref{cor}}
\corauth[cor]{Corresponding author.}
\ead{livio.gibelli@polimi.it}

\address{Politecnico di Milano,
         Dipartimento di Matematica,
         Piazza Leonardo da Vinci 32,
         20133 Milano, Italy}

\begin{abstract}
We show how to accelerate the direct solution of the Boltzmann equation 
using Graphics Processing Units (GPUs). 
In order to fully exploit the computational power of the GPU,
we choose a method of solution which combines a finite difference 
discretization of the free-streaming term with a Monte Carlo evaluation of the 
collision integral.
The efficiency of the code is demonstrated by solving the two-dimensional 
driven cavity flow.
Computational results show that it is possible to cut
down the computing time of the sequential code of two order of magnitudes.
This makes the proposed method of solution a viable alternative to particle 
simulations for studying unsteady low Mach number flows. 
\end{abstract}


\end{frontmatter}

\section{Introduction}

Non-equilibrium gas flows are met in several different physical situations
ranging from the re-entry of spacecraft in upper planetary atmospheres to 
fluid-structure interaction in small-scale devices \cite{c88,b94}.
The correct description of nonequilibrium effects requires replacing the 
traditional hydrodynamic equations with the Boltzmann equation which, 
in the absence of assigned external force fields, reads 
\begin{equation}
\label{eq:BE}
\frac{\partial f }{\partial t}+\bv \cdot \nabla_{\br}f = \mathcal{C}(f,f) 
\end{equation}
In Eq.~(\ref{eq:BE}), the distribution function $f(\br,\bv|t)$ is the atomic
number density at the single atom phase space point $(\br,\bv)$ at time $t$.
The symbols $\br$ and $\bv$ denote atom position and velocity, respectively.
The left hand side of Eq.~\eqref{eq:BE} represents the rate of change of $f$ 
due to the indipendent motion of gas atoms. Effects of collisions are 
accounted for by the source term $\mathcal{C}(f,f)$ which is a non-linear 
functional of $f$
whose precise structure depends on the assumed atomic interaction forces.
Obtaining numerical solutions of Eq.~(\ref{eq:BE}) for realistic flow
conditions is a challenging task because it has the form of a non-linear
integro-differential equation in which the unknown function, $f$, depends
on seven variables.
Numerical methods used to solve Eq.~\eqref{eq:BE} can be roughly divided 
into three groups:
\begin{itemize} 
\item[(a)] Particle methods
\item[(b)] Semi-regular methods
\item[(c)] Regular methods
\end{itemize}   
Methods in group (a) originate from the Direct Simulation Monte Carlo (DSMC) 
scheme proposed by G.A. Bird \cite{b94}. They are by far the most popular and 
widely used simulation methods in rarefied gas dynamics. The distribution 
function is represented by a number of mathematical particles which move in the
computational domain and collide according to stochastic rules derived from 
Boltzmann equation. Macroscopic flow properties are usually 
obtained by time averaging particle properties. If the averaging time is long 
enough, then accurate flow simulations can be obtained by a relatively small 
number of particles. The method can be easily extended to deal with mixtures 
of chemically reacting polyatomic species \cite{b94} and to dense fluids 
\cite{fgl05}. Although DSMC (in its traditional implementation) is to be 
recommended in simulating most of rarefied gas flows, it is not well suited to 
the simulation of low Mach number or unsteady flows. Attempts have been made to
extend DSMC in order to improve its capability to capture the small deviations 
from the equilibrium condition met in low Mach number flows \cite{hh07,w08}.
However, in simulating high frequency unsteady flows, typical of microfluidics 
application to MEMS, the possibility of time averaging is lost or 
reduced. Acceptable accuracy can then be achieved by increasing the number of 
simulation particles or superposing several flow snapshots obtained from 
statistically independent simulations of the same flow; in both cases the 
computing effort is considerably increased. \\
Methods in groups (b) and (c) adopt similar strategies in discretizing the 
distribution function on a regular grid in the phase space and in using finite 
difference schemes to approximate the streaming term. However, they differ in 
the way the collision integral is evaluated.
In semi-regular methods $\mathcal{C}(f,f)$ is computed by Monte Carlo or 
quasi Monte Carlo quadrature methods \cite{f91,t05} whereas deterministic 
integration schemes are used in regular methods \cite{a01}. Whatever method is 
chosen to compute the collision term, the adoption of a grid in the phase space
considerably limits the applicability of methods (b) and (c) to problems where 
particular symmetries reduce the number of spatial and velocity variables. As a
matter of fact, a spatially three-dimensional problem would require a memory 
demanding six-dimensional phase space grid. Extensions to polyatomic gases are 
possible \cite{f07} but the necessity to store additional variables associated 
with internal degrees of freedom further limits the applications to 
multi-dimensional flows.
Therefore, until now the direct solution of the Boltzmann equation by 
semi-regular or regular methods has not been considered a viable alternative 
to DSMC for simulating realistic flows, not even for low speed and/or unsteady
flows.
The availability of low cost Graphics Processing Units (GPUs) has changed the 
situation.
Although GPUs have been originally developed for graphics applications, 
they have been increasingly used to do general  purpose scientific and 
engineering computing \cite{mbkj09,jk10}.
Mapping efficiently an algorithm on the 
SIMD-like architecture of the GPUs, however, is a difficult task which 
often requires the algorithm to be revised or even redesigned to both
balance the hardware structure benefits and meet the implementation 
requirements. For instance, preliminary tests, performed within the framework
of the research work described here, have shown that the standard form of DSMC
is not efficiently ported on GPU's because of their SIMD-like architecture. 
On the other hand, we have shown in Ref.~\cite{fgg10} that a regular method of solution of the BGKW kinetic model equation is ideally suited for GPUs. 
The main aim of the present paper is to translate efficiently a 
semi-regular method of solution of the full non-linear Boltzmann equation into a parallel code to be executed on a GPU.
The efficiency of the algorithm is assessed by solving the classical 
two-dimensional driven cavity flow.
It is shown that it is possible to cut down the computing time of the
sequential code of two order of magnitudes. 
This paper is organized as follows. Sections 2 and 3 are devoted to a concise
description of the mathematical model and the adopted numerical method. 
In Section 4 the key aspects of the GPU
hardware architecture and \cuda programming model are briefly described
and implementation details are provided.
Sections 5 is devoted to the description of the test problem and the 
discussion of the results. Concluding remarks are presented in Section 6.

\section{Mathematical Formulation}\label{sec:MathForm}
The hard-sphere model is a good approximation for simple fluids, that is
fluids whose properties are largely determined by harshly repulsive short 
range forces. The hard-sphere Boltzmann collision integral reads

\begin{equation}
\label{collision_integral}
\mathcal{C}(f,f)
= \frac{\sigma^2}{2} \int
\left( f^{*}f_{1}^{*}- f f_{1} \right) |\bkhat \cdot \bv_r | 
d \bv_1 d^2 \bkhat
\end{equation}
In Eq.~\eqref{collision_integral},
$\sigma$ is the hard sphere diameter, $\bv_r=\bv-\bv_{1}$ is the relative
velocity between two colliding atoms and  
$f^{*}=f(\br,\bv^{*}|t), f_{1}^{*}=f(\br,\bv_{1}^{*}|t)$,
$f_{1}=f(\br,\bv_{1}|t)$.
Here and in the remainder of the paper, integration extends over the whole velocity space. Similarly, the solid 
angle integration is over the surface of the unit sphere, whose points are associated with the unit vector $\bkhat$.
The pre-collisional velocities, $\left(\bv^{*},\bv_{1}^{*}\right)$, are 
obtained from the post-collision velocities,
$\left(\bv,\bv_{1}\right)$, and the unit vector on the sphere, $\bkhat$,  
by the relationships
\begin{eqnarray}
\bv^* &=& \bv + \left( \bv_r\cdot \bkhat \right) \bkhat \\
\bv_{1}^* &=& \bv_{1} - \left( \bv_r \cdot \bkhat \right) \bkhat
\end{eqnarray}
In view of the applications to the study of low Mach flows, 
Refs. \cite{hh07,bh08} will be followed to rewrite Eqs. 
(\ref{eq:BE}) and (\ref{collision_integral}) in terms of the
deviational part of the distribution function, $h(\br,\bv|t)$, defined as
\begin{equation}
\label{definition}
f = \Phi_{0} \left( 1+\epsilon h \right)
\end{equation}
where $\epsilon$ is a parameter that measures the deviation from equilibrium 
conditions and
$\Phi_{0}(\br,\bv)$ is the Maxwellian at equilibrium 
with uniform and constant density
$n_{0}$ and temperature $T_{0}$, i.e.,
 \begin{equation}
\label{eq:MaxwellDist}
\Phi_{0} = \frac{n_{0}}{\left(2\pi R T_{0} \right)^{3/2}}
\exp\left(-\frac{\bv^2}{2RT_{0}}\right)
\end{equation}
The physical rationale behind this formulation is a proper rescaling
of the (small) deviation from equilibrium to reduce the
variance in the Monte Carlo evaluation of the collision integral and thus
to capture arbitrarily small deviations from equilibrium with a computational cost which is independent of 
the magnitude of the deviation.
By substituting Eq.~\eqref{definition} into Eq.~\eqref{eq:BE}, we obtain
\begin{equation}
\label{eq:devBoltz}
\frac{\partial h}{\partial t}+
\bv \cdot \nabla_{\br}h =\mathcal{Q}(h,h)
\end{equation}
where, by using the property 
$\Phi_{0}^{*}\,\Phi_{01}^{*}=\Phi_{0}\,\Phi_{01}$,
the collision integral takes the form
\begin{equation}
\label{eq:devIntColl}
\mathcal{Q}(h,h) =
\frac{\sigma^2}{2} \int
\Phi_0\,\Phi_{01}
\bigl[h^* + h_{1}^*-
       h   - h_{1}  +
       \epsilon \, (h^*h_{1}^* - h h_{1}) \bigr]
|\bkhat \cdot \bv_r| d\bv_1 d^2 \bkhat
\end{equation}
For later reference, we here report the expressions of dimensionless perturbed 
density, velocity, temperature and stress tensor
\begin{eqnarray}
\label{densita}
\rho &=& \frac{n-n_{0}}{n_{0}} \frac{1}{\epsilon} = \int \Phi_0 \,h \,d\bv \\
\label{velocita}
\bu & = & \frac{\bV}{\sqrt{2RT_0}}\frac{1}{\epsilon}= 
\frac{1}{1+\epsilon\,\rho}\int \Phi_0 \,h \, \bv \,d\bv \label{eq:vel}\\
\label{temperatura}
\theta &=& \frac{T-T_0}{T_0}\frac{1}{\epsilon} = \frac{1}{1+\epsilon\,\rho}
\Bigl( \frac{1}{3}\int \Phi_0 \,h \, \bv^2 \,d\bv
- \rho \Bigr) -\frac{\epsilon}{3}\bu^2\\
\label{sforzi}
\Pi_{ij} &=& \frac{p_{ij} - p_0\delta_{ij}}{p_0}\frac{1}{\epsilon} =
\int \Phi_0 \,h \, v_i v_j \, d\bv -
\epsilon\,u_i\,u_j -\epsilon^2\,\rho\,u_i\,u_j 
\end{eqnarray}
where $p_{0}=n_{0} R T_{0}$.
At the boundaries, Maxwell's completely diffuse boundary condition is assumed. Accordingly, the distribution function of atoms emerging from walls is given by the following expression

\begin{equation}
\label{eq:bcMax}
\Phi_0 +\epsilon \, \Phi_0 \, h = n_w \, \Phi_w \quad 
\quad \left( \bv-\bV_w \right) \cdot \boldsymbol{\hat n}>0
\end{equation}
In Eq.~(\ref{eq:bcMax}), $\boldsymbol{\hat n}$ is the inward normal and
$\Phi_w$ is the normalized wall Maxwellian distribution function 

\begin{equation}
\Phi_w(\br, \bv) = \frac{1}{(2\pi RT_{w})^{3/2}}
\exp{\left[ -\frac{(\bv-\bV_{w})^{2}}
                  {2RT_{w}} \right]}
\end{equation}
where $\bV_{w}$ the wall velocity and $T_w$ the wall
temperature. The wall density $n_{w}$ is
determined by imposing zero net mass flux at any boundary point
\begin{equation}
n_w \int_{\bc \cdot \boldsymbol{\hat n}>0} |\bc\cdot
\boldsymbol{\hat n}| \Phi_w d{\bv}= \int_{\bc\cdot
\boldsymbol{\hat n}<0} |\bc\cdot \boldsymbol{\hat n}|
\Phi_0 d{\bv} + \epsilon \int_{\bc\cdot \boldsymbol{\hat
n}<0} |\bc \cdot \boldsymbol{\hat n}| \Phi_0 \,h d{\bv}
\end{equation}
where $\bc=\bv-\bV_w$.
It is worth noticing that when the perturbation is
sufficiently small, i.e., $\epsilon \rightarrow 0$,
Eq.~(\ref{eq:devBoltz}) reduces to the linearized Boltzmann equation and
Eqs. (\ref{densita})-(\ref{sforzi}) to the linearized expression of the 
macroscopic quantities. The formulation in terms of the deviational part of
the distribution function, however, is not restricted to
a vanishing perturbation but it is valid in the non-linear case as well.

\section{Outline of the numerical method}
The method of solution adopted to solve Eq.~\eqref{eq:devBoltz} is 
a semi-regular method in which a finite difference discretization is used
to evaluate the free streaming term on the left hand side while the collision
integral on the right hand side is computed by a Monte Carlo technique.
The three-dimensional physical space is divided into
$N_r=N_x \times N_y \times N_z$ parallelepipedal cells. Likewise, the
three-dimensional velocity space is replaced by a parallelepipedal
box divided into $N_v=N_{v_x}\times N_{v_y}\times N_{v_z}$ cells.
The size and position of the ``velocity box'' in the velocity space have to
be properly chosen, in order to contain the significant part of
$h$ at any spatial position. The distribution
function is assumed to be constant within each cell of the
phase space. Hence, $h$ is represented by the array
$h_{\bi,\bj}(t)=h(x(i_x),y(i_y),z(i_z),v_x(j_x),v_y(j_y),v_z(j_z)|t)$;
$x(i_x),y(i_y),z(i_z)$ and $v_x(j_x),v_y(j_y),v_z(j_z)$ are the values of
the spatial coordinates and velocity components in the center
of the phase space cell corresponding to the indexes
$\bi=(i_x,i_y,i_z)$ and $\bj=(j_x,j_y,j_z)$.\\
The algorithm that advances
$h_{\bi,\bj}^{n}=h_{\bi,\bj}(t_{n})$ to
$h_{\bi,\bj}^{n+1}=h_{\bi,\bj}(t_{n}+\Delta t)$ is constructed
by time-splitting the evolution operator into a free streaming
step, in which the right hand side of Eq.~(\ref{eq:devBoltz}) is
neglected, and a purely collisional step, in which spatial
motion is frozen and only the effect of the collision operator is taken
into account. More precisely, the distribution function
$h_{\bi,\bj}^n$ is advanced to $h_{\bi,\bj}^{n+1}$ by computing
an intermediate value, $\tilde{h}_{\bi,\bj}^{n+1}$, from the free streaming 
equation
\begin{equation}
\frac{\partial h }{\partial t}+ \bv \cdot \nabla_{\br} h =0
\label{eq:freestreaming}
\end{equation}
When solving Eq.~\eqref{eq:freestreaming}, 
boundary conditions have to be taken into account.
Eq.~(\ref{eq:freestreaming}) is discretized by a simple first
order explicit upwind conservative scheme. 
For later reference, we here report the difference scheme in the 
two dimensional case with $v_{x}>0$ and $v_{y}>0$

\begin{equation}
\label{eq:streaming2D}
\tilde{h}^{n+1}_{i_x,i_y; \bj} = 
(1-\mbox{Cu}_x\;-\mbox{Cu}_y) h^{n}_{i_x,i_y; \bj}+
\mbox{Cu}_x \; h^{n}_{i_x-1,i_y; bj} + \mbox{Cu}_y \; h^{n}_{i_x,i_y-1; \bj}
\end{equation}
In Eq.~(\ref{eq:streaming2D}),
$\mbox{Cu}_x=v_x(j_x)\Delta t/\Delta x$ and $\mbox{Cu}_y=v_y(j_y)\Delta
t/\Delta y$ are the Courant numbers in the $x$ and $y$
directions, respectively. \\
After completing the
free streaming step, $h_{\bi,\bj}^{n+1}$ is obtained by solving
the homogeneous relaxation equation
\begin{equation}
 \frac{\partial h}{\partial t}=\mathcal{Q}(h,h)
\label{eq:homrel}
\end{equation}
where $\mathcal{Q}(h,h)$ is given by
Eq.~\eqref{eq:devIntColl}. 
In order to be solved, Eq.~(\ref{eq:homrel}) is first integrated
over the cell of the velocity space $C_{\bj}$
\begin{equation}
\label{eq:intSuCj}
 \frac{d N_{\bi,\bj}}{d t} = \int_{C_{\bj}} \mathcal{Q}(h,h) d \bv
\end{equation}
where $N_{\bi,\bj}$ represents the deviation of the number of particles 
with position $\br_{i}$ in 
the velocity cell centered around the velocity node $\bj$ with respect to its
mean value at equilibrium, i.e.,
$N_{\bi,\bj} \backsimeq \Delta \mathcal{V}_{\bj}\, \Phi_{0,\bj}
\,h_{\bi,\bj}$ with $\Delta \mathcal{V}_{\bj}$ the volume of the velocity
cell $C_{\bj}$.
The integral in Eq.~(\ref{eq:intSuCj}) is then
transformed into an integral extended to the whole velocity
domain $\mathcal{V}$ 
\begin{equation}
\label{integ}
\frac{d N_{\bi,\bj}}{d t} = \int_{\mathcal{V}} \chi_{\bj} 
                            \,\mathcal{Q}(h,h)\, d \bv
\end{equation}
where $\chi_{\bj}$ is the
characteristic function of the cell $C_{\bj}$
\begin{equation}
\chi_{\bj}(\bv)=
\begin{cases}
1 & \bv \in C_{\bj}\\
0 & \bv \notin C_{\bj}
\end{cases}
\end{equation}
Making use of some fundamental properties of the collision
integral \cite{c88}, Eq.~(\ref{integ}) can be written in
the following form
\begin{multline}
\label{eq:coll8Dim} 
\frac{dN_{\bi,\bj}}{dt}  = \frac{d^2}{4}
\int d \bv \, d\bv_{1}\;
\Phi_{0}(\bv) \, \Phi_{0}(\bv_{1}) \int_{-1}^{1} dk_{z}
\int_{0}^{2\pi} d\phi\\
\left[ \chi_{\bj}(\bv^{*}) + \chi_{\bj}(\bv^{*}_{1}) -
       \chi_{\bj}(\bv) - \chi_{\bj} (\bv_{1}) \right]
\left[ h(\bv) + h(\bv_{1}) + \epsilon h(\bv) h(\bv_{1}) \right] 
|\bkhat \cdot \bv_r|
\end{multline}
The eight-fold integral in Eq.~(\ref{eq:coll8Dim}) is
calculated by a Monte Carlo integration method, since a regular
quadrature formula would be too demanding in term of computing
time. The advantage of writing the rate of change of $N_{\bi,
\bj}$ in the above form is that the gaussian distribution
function $\Phi_0$ may be considered a probability density
function from which the velocity points are drawn to estimate
the collision integral with lower variance. 
The Monte Carlo estimate of the integral on the right hand side of 
Eq.~\eqref{eq:coll8Dim} gives
\begin{multline}
\label{eq:monteCarloInt} 
\frac{dN_{\bi, \bj}}{d t} =
\frac{n_{0}^{2} d^2 \pi}{N_{t}} \sum_{l=1}^{N_{t}} \left[
\chi_{\bj}(\bv^{*}_{l}) + \chi_{\bj}(\bv_{1l}^{*}) -
\chi_{\bj}(\bv_{l}) - \chi_{\bj}(\bv_{1l})
                           \right] \\
\left[ h(\bv_{l}) + h(\bv_{1l}) + \epsilon \,
h(\bv_{l})\,h(\bv_{1l}) \right] |\bkhat \cdot \bv_{r}|
\end{multline}
where $N_t$ is the number of velocity samples \cite{f91}.
It is worth noticing that the same set of collisions can be used to 
evaluate the collision integral at different space locations. 
Once the collision integral have been evaluated, the solution
is advanced from the $n$-th time level to the next according to
the explicit scheme
\begin{equation}
\label{collision_step}
h_{\bi,\bj}^{n+1} = \tilde{h}_{\bi,\bj}^{n+1} +
\frac{1}{\Delta \mathcal{V}_{\bj}\, \Phi_{0,\bj}}
\frac{d N_{\bi,\bj}}{dt} \Delta t
\end{equation}
In Eq.~\eqref{collision_step}, $d N_{\bi,\bj}/dt$ is given by 
Eq.~\eqref{eq:monteCarloInt} with $h$ the deviational part of the 
distribution function at the end of the streaming step, that is 
$\tilde{h}^{n+1}_{\bi,\bj}$.
Although memory demanding, the method outlined above produces
accurate approximations of the solution which do not require
time averaging to provide smooth macroscopic fields. A drawback
of the technique is that, due to the discretization in the
velocity space, momentum and energy are not exactly conserved.
The numerical error is usually small but tends to accumulate
during the time evolution of the distribution function. The
correction procedure proposed in Ref.~\cite{at80} has been
adopted to overcome this difficulty. At each time step the full
distribution function is corrected in the following way
\begin{equation}
\label{eq:rinorm_a}
\Phi_{0,\bj}\bigl(1+\epsilon \, h^{n+1}_{\bi,\bj}\bigr)
= \Phi_{0,\bj}\bigl(1 + \epsilon \, \tilde{h}^{n+1}_{\bi,\bj}\bigr)\left[
1+A+\bB \cdot\bv+C\bv^{2}\right]
\end{equation}
where the constants $A,\bB$ and $C$ are determined
from the conditions
\begin{equation}
\label{eq:correzione}
\int \psi(\bv)\;\Phi_{0}(\bv)\,h^{n+1}(\bv)\,d \bv =
\int \psi(\bv)\;\Phi_{0}(\bv)\, \tilde{h}^{n+1}(\bv)\,d \bv
\end{equation}
being $\psi(\bv)=1,\bv,\bv^{2}$.
The correction procedure given by Eq.~\eqref{eq:rinorm_a} involves
the full distribution function and not only its deviational part in order 
the linear system (\ref{eq:correzione}) to be well conditioned.

\section{GPU and \cuda} 

\subsection{Overview of GPU computing}
\nvidia
GPU is built around a fully programmable processors array organized
into a number of multiprocessors with a SIMD-like architecture, 
i.e. at any given clock cycle, each core of the multiprocessor 
executes the same instruction but operates on different data.
\cuda
is the high level programming language 
specifically created for developing applications on this platform
\cite{n08}. \\
A \cuda 
program is organized into a serial program
which runs on the host CPU and one or more
kernels which define the computation to be performed in parallel by a massive 
number of threads. 
Threads are organized into a three-level hierarchy.
At the highest level, all threads form a grid; they all execute the same
kernel function. Each grid consists of many different blocks which contain
the same number of threads. 
A single multiprocessor can manage a number of blocks concurrently 
up to the resource limits. 
Blocks are independent, meaning that a kernel must 
execute correctly no matter the order in which blocks are run.
A multiprocessor executes a group of threads beloging to the active block,
called warp.
All threads of a warp execute the same instruction but operate on different 
data. 
If a kernel contains a branch and threads of the same warp follow different
paths, then the different paths are executed sequentially (warp divergence) 
and the total run time is the sum of all the branches. 
Divergence and re-convergence are managed in hardware but may have a serious 
impact on performances.
When the instruction has been executed, the multiprocessor
moves to another warp. In this manner 
the execution of threads is interleaved rather than simultaneous. \\
Each multiprocessor has a number of registers which are dynamically 
partitioned among the threads running on it. Registers are memory spaces 
that are readable and writable only by the thread to which they are 
assigned. Threads of a single block are allowed to synchronize with each other 
and are available to share data through a high-speed shared memory. 
Threads from different blocks in the same grid may coordinate only via 
operations in a slower global memory space which is readable and writable 
by all threads in a kernel as well as by the host. 
Shared memory can be accessed by threads within a block as quickly as
accessing registers. On the contrary, I/O operations involving global memory are particularly
expensive, unless access is coalesced \cite{n08}.
Because of the interleaved warp execution,
memory access latency is partially hidden, i.e.,
threads which have read their data can be performing
computations while other warps running on the same multiprocessor are
waiting for their data to come in from global memory. 
Note, however, that GPU global memory is still ten time faster than the main
memory of recent CPUs.\\ 
Code optimization is a delicate task.
In general, applications which require many arithmetic operations between
memory read/write, and which minimize the number of out-of-order memory
access, tend to perform better. Number of blocks 
and number of threads per block have to be chosen carefully.
There should be at least as many blocks as multiprocessors in the
device. Running only one block per multiprocessor can force the multiprocessor 
to idle during thread synchronization and device memory reads.
By increasing the number of blocks, on the other hand, 
the amount of available shared memory for each block diminishes.
Allocating more threads per block is better for efficient time slicing,
but the more threads per block, the fewer registers are available per thread.

\subsection{\cuda implementation}
The code to numerically solve
Eq.~(\ref{eq:devBoltz}) is organized into a host program,
which deals with all memory management and other setup tasks,
and three kernels running on the GPU which perform the
streaming and the collision steps. 
In the following, we report and discuss the pseudo-codes of 
each kernel.
Because of their different impact on the code performance, we
distinguish the slow global memory reads, 
$\Leftarrow$, and writes, $\Rightarrow$, 
from the fast reads, $\leftarrow$, and writes, $\rightarrow$,
from local registers and shared memory.

\noindent
Algorithm \eqref{alg:streaming2D} reports the pseudocode
of the two dimensional streaming kernel. The one-dimensional case
has been discussed in Ref. \cite{fgg10} whereas the extension to 
three-dimensional geometries is straightforward. 
Moreover, for clarity of presentation, the pseudo-code of the streaming kernel 
refers to one cell of the velocity space with $v_{x}>0$ and $v_{y}>0$.
The other cases can be handled analogously.
As shown by Eq.~(\ref{eq:streaming2D}), for each cell of the
velocity domain, the streaming step involves the distribution function
evaluated at different space locations.
Similarly to the one dimensional case, the key performance
enhancing strategy is to allow threads to cooperate in the shared memory. 
In order to fit into the device's resources, blocks are composed
by a two dimensional grid of threads with dimension $B_{x} \times
B_{y}$ having each thread associated with one cell of the physical space.
When a block become active, each thread loads one element of
the distribution function from global memory, stores it into shared
memory (line 5), updates its value according to Eqs. (\ref{eq:streaming2D})
(line 21) and then saves it back to the global memory (line 22). 
This procedure is repeated sequentially 
$\left( N_{x}/B_{x}-1 \right) \times \left(N_{y}/B_{y}-1\right) $ times.
To ensure non-overlapping access, threads are synchronized at the onset of
writing to the global memory (lines 20).
In order to obtain a coalesced access to the global memory, values of the 
discretized distribution function of cells which are adjacent in 
the physical space are stored in contiguous memory locations.
However, not all the threads in a block can read data in a coalescent manner. 
In fact, in order to update $h^{n}_{\bi,\bj}$ the values of the distribution 
function of two upwind neighboring nodes, often referred to as ``halo'' nodes
\cite{M09}, are required.
The halos on one physical direction 
can be read in with coalesced access (line 13-19) while the others
have to be read in with non-coalesced access (line 6-12).
Threads which update boundary points perform calculations which are 
slightly different to account for the incoming Maxwellian flux from the
boundaries of the domain (lines 8 and 15). \\
The relaxation step is organized into two kernels whose pseudo-codes
are listed in Algorithms (\ref{alg:sampling}) and (\ref{alg:collision}).
The first kernel computes the sequence of $N_{t}$ collisions used in the
Monte Carlo evaluation of the collision integral. 
The second kernel updates the discretized distribution function, executes the 
correction procedure and computes the macroscopic quantities of interest as
well. \\
Algorithm~\eqref{alg:sampling} reports the pseudo-code of the sampling
kernel. Here, there are as
many threads as the number of the collision samples, $N_t$. Firstly,
each thread generates the pre-collisional velocities $\bv$ and
$\bv_{1}$ by sampling the maxwellian distribution
function with the Box-Muller algorithm (line 1-2) and the
unit vector $\bkhat$ by sampling the uniform distribution on
the unit sphere (line 3). Afterwards, the post-collisional velocities are
evaluated (line 4-6) and the index of the velocity cells to which they belong
are calculated and stored in the vectors $I$, $I_{1}$, $I^{*}$
and $I_{1}^{*}$ defined in the global memory (lines 7-10). 
Finally, for each
velocity cell, the values to be added and subtracted to these velocity cells
are calculated (lines 11-14) and  stored (lines 15-18) in the
vectors $C$, $C_1$, $C^{*}$ and $C_{1}^{*}$ defined in the global
memory. It is important to note that in order to maximize the
performance all the accesses to the global memory are done in a
coalesced manner \cite{n08}. \\
To update the discretized distribution function in a cell of the physical 
space according to Eq.~(\ref{collision_step}),  
no information from nearby space cells is required.
This naturally fits for the GPU, where one may define as many threads
as the number of cells in the physical space. Moreover,
by having one thread for each cell of the physical space,
potentially dangerous conflicts between threads are avoided and the 
accesses to the global memory may be coalesced. 
Firstly, each thread updates the discretized
distribution function according to Eq.~\eqref{eq:monteCarloInt},
then executes the correction procedure to enforce the
conservation of momentum and energy, Eq.~\eqref{eq:rinorm_a},
and finally compute the macroscopic quantities of interest.
Algorithm \eqref{alg:collision} shows the pseudo-code of the relaxation kernel.
The main part of the algorithm is in between the
lines 7 and 21 where the collision integral is evaluated
according to the Monte Carlo method,
Eq.~\eqref{eq:monteCarloInt}. Lines from 1 to 6 and  
from 22 to 30 evaluate the required moments of the distribution function
before and after the collision step, respectively. These
moments are used in Eq.~\eqref{eq:correzione} to obtain the
constants $A, \boldsymbol{B}$ and $C$.
The last loop over the velocity space (lines 32-39) corrects
the distribution function according to
Eq.~\eqref{eq:rinorm_a} and compute the macroscopic
quantities of interest. \\
The computations that are shown below, have been performed on a 
commercially available GPU GeForce GTX 260 produced by 
$\mbox{NVIDIA}^\circledR$ using $\mbox{CUDA}^{\mbox{\tiny TM}}$ version 2.0.
The GTX 260 GPU model consists of 24 streaming multiprocessors with 8 streaming processors (SP) each for a total of 192 units. Each SP is clocked at
1.242 GHz and performs up to 3 floating point operation (FLOP) per clock cycle,
yielding a peak theoretical performance of 715.4 GFLOPs
($192 \times 1.242 \times 3$). Each group of SP shares one 16 kB of fast
per-block shared memory while the GPU has 896 MB of device memory with a
memory bandwidth of 111.9 GB/s.
The graphic processing unit has been hosted by a personal computer equipped
with 4 GB of main memory and an $\mbox{Intel}^\circledR$
Core Duo Quad Q9300 CPU, running at 2.5 GHz.
The host machine has also been used to run the sequential version of the
program to obtain the speed-up data.
The host code has been compiled using the gcc/g++ compiler
with optimization option ``-03''.

\section{Test case: driven cavity flow}

\subsection{Formulation of the problem}
The driven cavity flow is a classical benchmark problem. 
In spite of its simple geometry, in fact, it contains most of the features
of more complicated problems described by kinetic equations \cite{VVS08}.
In the following, we restrict to the spatially two-dimensional case. We
thus consider a monatomic rarefied gas contained in a square enclosure with 
length $L$, i.e., $\br \in [-L/2,L/2]\times[0,L]$.
All the walls are kept at uniform and constant temperature
$T_0$. 
Initially, the gas is supposed to be in uniform equilibrium with density 
$n_0$ and temperature $T_0$.
The flow is driven by the translation of the lid of the 
cavity with constant velocity $V_w$. 
We describe the dynamics of the gas by Eq.~\eqref{eq:devBoltz} and assume
that atoms which strike the walls are
re-emitted according to the Maxwell's scattering kernel with
complete accomodation, Eq.~\eqref{eq:bcMax}. \\
As characteristic length
we choose $\lambda_{0}=\mu_{0}/p_{0} (2 RT_{0})^{1/2}$,  
with $\mu_{0}$ the viscosity of the hard sphere gas \cite{cc90}.
Likewise the characteristic time is given by $\lambda_{0}/V_{0}$,
where $V_{0}=(2RT_{0})^{1/2}$.
The cavity flow problem has been solved for three values of the rarefaction 
parameter $\delta=L/\lambda_{0}=0.1,1,10$, 
being $\delta$ proportional to the reciprocal of the Knudsen number.
Since the proposed method of solution is particularly effective in capturing 
small deviations from equilibrium,
we set the lid velocity to $V_{w}/V_{0}=0.01$.
The computations described in below, hence, refer to very low Mach number
driven cavity flows. 
The square cavity has been divided into $N_{r}=N_{x}\times N_{y}$ uniform
square cells, with $N_{x}=N_{y}$.
Likewise the velocity space has been divided into 
$N_v=N_{v_{x}}\times N_{v_{y}}\times N_{v_{z}}$ 
with $N_{v_{x}}=N_{v_{y}}=N_{v_{z}}$.
Since the deviation form equilibrium is supposed to be small, a cubic velocity
space has been constructed by distributing 
the velocity nodes 
along each velocity component in the interval $[-3 V_0,3 V_0]$.
In order to achieve a faster convergence of the solutions in the velocity
space, the lengths of the cells are uniformly stretched with 
a progression ratio $r_v$, being
the smaller cells located at the origin of the velocity space.
More precisely, it has been chosen
$N_v=8000$ and $r_{v}=0.945$ for $\delta=0.1$ and
$\delta=1$ 
whereas 
$N_v=5832$ and $r_{v}=0.970$ for $\delta=10$ 
The number of collisions used in the Monte
Carlo evaluation of the collision integral have been varied with the 
rarefaction parameter. In particular, it has been set
$N_{t}=1024$ for $\delta=0.1$, 
$N_{t}=6144$ for $\delta=1$ and 
$N_{t}=8192$ for $\delta=10$. 
Finally, the time step has been determined by setting the maximum Courant
number to $0.5$.

\subsection{Results and discussion}
In this section, we first carry out a convergence analysis of the method
in the physical space
and then we investigate the parallel performances of the code.\\
In order to establish the convergence rate we compute two global flowfield 
properties, namely the mean dimensionless shear stress on the moving wall, $D$,
and the dimensionless flow rate of the main vortex, $G$. The two quantities 
are defined as 
\begin{equation}
D=\frac{1}{\delta}\int_0^{\delta} \Pi_{xy}(x,\delta|t)\,dx, \qquad
G=\frac{1}{\delta}\int_0^{\delta} |u_x(\delta/2,y|t)|\,dy
\end{equation}
The absolute relative error in the long-term mean values of $D$ and $G$ are 
shown in Figs~\ref{fig:e_D_G}a and \ref{fig:e_D_G}b, respectively, 
versus the spatial grid size, $h/\delta=1/\sqrt{N_{r}}$, 
and for $\delta=0.1$ (circles),
$\delta=1$ (squares) and $\delta=10$ (triangles).
The exact values of $D$ and $G$, which are referred to as $D_{e}$ and 
$G_{e}$, have been extrapolated from the linear fit of the results 
when $h\rightarrow 0$. 
The linear behaviour of the absolute relative errors demonstrates that
the results are in the asymptotic range of convergence and the method is first 
order accurate \cite{s06}.
The finest physical grid size provides predictions which are accurate only
within few percent.
More precisely, the largest error in 
$D$ is of the order of $4\%$ and is attained at $\delta=10$ whereas the one
in $G$ is  $2\%$ at $\delta=0.1$.
The error is mainly due to the physical and velocity discretizations.  
As a matter of fact the statistical error associated with the finite sample 
size used in the Monte Carlo evaluation of the collision integral does not 
affect the results significantly. The standard deviation of $D$ and $G$ with 
respect to their averaged values in stationary conditions, in fact, is
negligible small. For instance, the standard deviation of $D$ at $\delta=1$ 
from its long-term mean value is less than 0.05\%. 
The grid resolution in the physical and the velocity domains
are the more accurate discretization 
compatible with the GPU global memory constraint, i.e.,
$N_r=25600$, $N_v=8000$ for $\delta=0.1$ and $\delta=1$ whereas 
$N_r=36864$, $N_v=5832$ for $\delta=10$.
Therefore, in order to improve the accuracy of the numerical solutions, 
we have adopted a nonuniform grid in the physical space.
The lengths of the physical cells are uniformly stretched 
with progression ratios that depend on the rarefaction parameter,
$r_x=0.990$, $r_y=0.995$ for $\delta=10$ and $r_x=0.996$, $r_y=0.998$ 
otherwise. 
The smaller cells are located
close to the upper corners of the cavity where severe gradients are 
anticipated. 
All the results which follow have been obtained with these discretizations.
Table \ref{tab:DeG} compares the predictions of the long-term mean values of
$D$ and $G$ with the extrapolated exact values, $D_e$ and $G_e$, and the
results reported in Ref.~\cite{VVS08} where the linearized
BGKW equation has been solved with a discrete velocity method.
The accuracy of the numerical solution of the non-linear Boltzmann equation 
can be estimated to be within $2$\%.
The agreement with the BGKW results is good, which is not unexpected. 
Since the velocity of the lid is small, in fact, the gas is in a weakly
nonequilibrium state and the solution of the
non-linear Boltzmann equation approaches the solution of the
linearized BGKW equation. 
Figures \ref{fig:profili}a and \ref{fig:profili}b show the profiles of the 
dimensionless
horizontal component of the velocity, $V_{x}/V_{w}$, along the vertical line 
crossing the center of the cavity and
the dimensionless vertical component of the velocity, $V_{y}/V_{w}$,
along the horizontal line
crossing the center of the main vortex which forms in the cavity,
respectively. Lines are the solutions of the non-linear 
Boltzmann equation, whereas symbols are the
results reported in Ref.~\cite{VVS08}.
The results refer to two different values of the rarefaction
parameters, $\delta=0.1$ (dashed lines and squares) and
$\delta=10$ (solid lines and circles). The agreement is very good.\\
Although the convergence analysis has been performed by referring to quantities
evaluated in stationary conditions, it is important to point out that the
proposed method of solution provides accurate results in the unsteady regime
as well.
As an example, Figure \ref{fig:dragVsTime} shows the evolution of $D$ during
the simulation for $\delta=10$. 
Such a result would be difficult to obtain 
with a particle method where computationally expensive ensemble averanging are 
needed to provide smooth macroscopic fields.

\noindent
The performance of the GPU implementation is compared against
the single-threaded version running on the CPU by computing the
speed-up factor {$S=T_{\mbox{\tiny CPU}}/T_{\mbox{\tiny GPU}}$,
where $T_{\mbox{\tiny CPU}}$ and $T_{\mbox{\tiny GPU}}$ are the
times used by the CPU and GPU, respectively}. Times are
measured after initial setup, and do not include the time
required to transfer data between the disjoint CPU and GPU
memory spaces.\\
We analyse separately the streaming and the collision step, the latter
comprising both the sampling and the collision kernel.  
Figure \ref{fig:speedUp} reports the obtained speed-up data
as a function of the number of spatial grid points $N_{r}$ for $\delta=1$. 
The speed-up grows
rapidly with $N_{r}$ and than levels up at about 450 if $N_{r}$
approximately exceeds $10^4$. This behavior is the result of
the parallel set up of the collision step in $N_{r}$
independent threads one for each cell of the physical space. As
discussed below, the collision step absorbs most of the
computational resources and its execution strongly affects the
overall performances. As shown by the speed-up curve, the GPU
power is not fully exploited till the number of concurrent
threads reaches a threshold. Beyond, the speed-up saturates and
the computing time approximately behaves as a linear function
of $N_{r}$. This behaviour is similar to the one reported in 
Refs.~\cite{eld08,alt08}.
Figure~\ref{fig:times} shows the relative time which is spent on 
the streaming step (dark bar) and on the collision step
(light bar) as well as the total execution time in seconds (numbers over  
the bars) for $\delta=1$. 
As expected, the collision step is more time consuming than the streaming
step which takes at most 36\% of overall computing time. \\
A strongly simplified evaluation of ideal performances of the
streaming and collision step can be carried out as follows.
A single application of the upwind scheme requires the execution of 11
floating point operations and 2.3 accesses to the global
memory. The GPU delivers 715.4 GFLOPs but
the transfer rate to/from the main memory is limited to 111.9
GB/s. Since in the case of the streaming step the ratio of number of
floating point operations to {the number of bytes accessed} is
low ($11:9.2$), it is reasonable to obtain the number of
floating point operation per second from the transfer rate
alone. Hence, the ideal number of GFLOPs can be obtained by
assuming that 11 floating point operations will be executed in
the time required to transfer 9.2 bytes from the main memory.
Accordingly, this simple argument yields an ideal performance of the
streaming step
of 133 GFLOPs. A similar performance analysis can be applied to the collision 
step which encompasses both the sampling and 
collision kernel. In order to update the distribution
function and compute the macroscopic quantities of interest,
the number of FLOPs that the GPU executes at any given time
step and for each cell in the physical space is the sum of two
contributions. The first is proportional to the number of
velocity cells, $N_{v}$, and the second one is
proportional to the number of collisions used to evaluate the
collision integral, $N_t$. The resulting number of FLOPs for each time
step are of the order of $N_r (80 N_{v} + 12 N_t)$.
Likewise the number of bytes accesses to the global memory per time step is
of the order of $N_r (8 N_v + 64 N_t)$
Arguments similar to those above lead to estimate an ideal performace of
the collision step of about 
174.7 GFLOPs.\\
Timing the execution of the separate kernels and counting the
number of associated floating point operations provides the
real performance. The results are reported in
Fig.~\ref{fig:GflopDriCav} where GFLOPs are shown as a
function of the number of grid points, $N_{r}$. 
Solid line with circles, dashed line with
squares and dot-dashed line with triangles are the measured
performances of the streaming step, the collision step and the overall code.
respectively. It is possible to note that the performance of
the streaming step grows with $N_{r}$ and quickly levels at
about $30$ GFLOPs, approximately one fourth of the estimated
ideal performance. The difference can be justified by observing
that the real \cuda implementation of the finite difference
scheme is not free from thread divergence and
ancillary tasks whose effects can be evaluated with difficulty \cite{M09}.
The collision step performance closely patterns the speed-up
behavior, that is it rapidly grows in the range $N_{r}<10^4$ and
then levels up at about $140$ GFLOP/s, reasonably close to the 
theoretical estimate. The 
collision kernel performs better than the streaming kernel due to
its higher {FLOP to memory operation ratio} which, in turn, allows a more
efficient use of GPU computing power. The absence of thread
divergence is also a feature which positively affects
performances.

\section{Conclusions}

In this paper we have assessed the possibility of exploiting
the computational power of modern GPUs to simulate nonequilibrium
rarefied gas flows.
The full nonlinear Boltzmann equation has been solved by means of a 
semi-regular method which combines a finite difference 
discretization of the free-streaming term with a Monte Carlo evaluation of the 
collision integral. This method of solution is ideally suited for the
SIMD-like architecture provided by the commercially available GPUs.
The two dimensional driven cavity flow has been used as a benchmark problem.
The results lead to concluding that the porting of the
sequential code onto GPUs allows a reduction of the computing
time of two orders of magnitude, being the observed speed-up as high as 400. 
Although the test problem examined here has clearly shown that the size of
physical memory is the main obstacle toward the application to
complex two or three-dimensional flows, the numerical method described above
can be reformulated as a less memory demanding particle scheme in many ways. 
Hence, the present work and results
are a first step toward the construction of a more flexible and efficient
method for the numerical solution of kinetic equations.

\section*{Acknowledgment}
Support received from {\bf Fondazione Cariplo}
within the framework of project \emph{``Fenomeni dissipativi e di rottura in
micro e nano sistemi elettromeccanici''}, and
{\bf Galileo Programme} of Universit\`a Italo-Francese within
the framework of project MONUMENT (MOdellizzazione NUmerica in MEms e
NanoTecnologie) is gratefully acknowledged.
The authors wish to thank Professor Dimitris Valougeorgis for providing his
numerical results.

\newpage

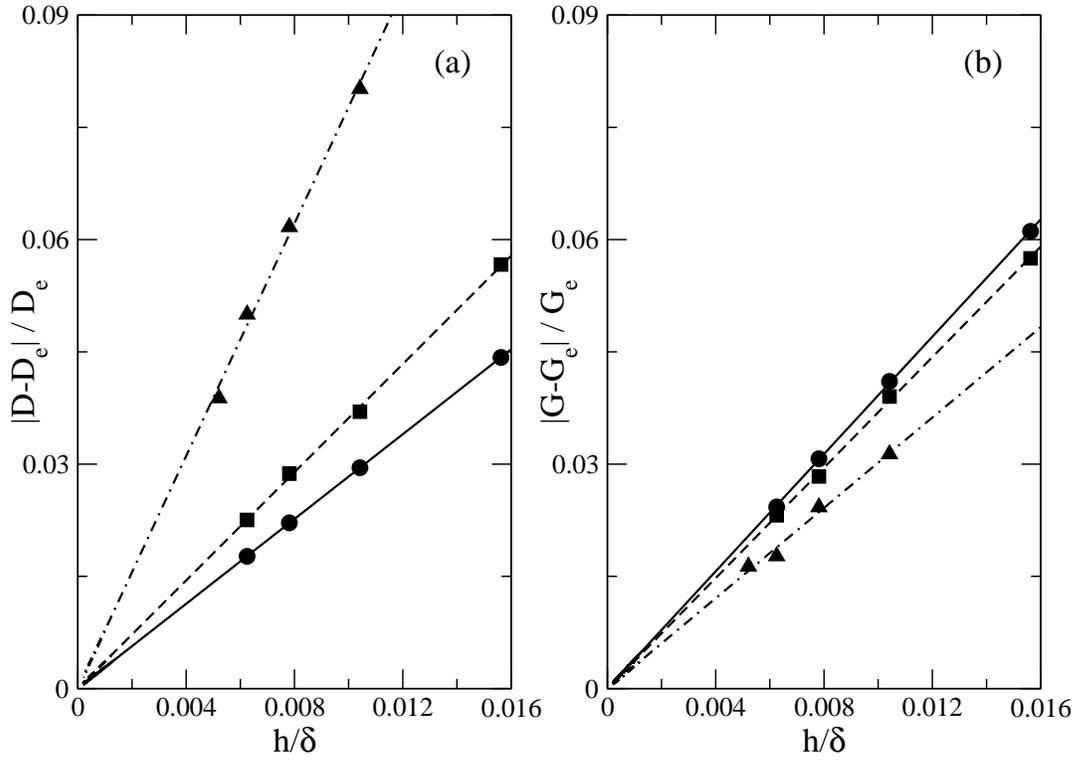
\begin{figure}[h!]
\begin{center}
\epsfig{file=Fig1.eps,height=10cm}
\caption{}
\label{fig:e_D_G}
\end{center}
\end{figure}

\noindent
Captions to Figures:

\vspace{1cm}

\noindent
Fig. \ref{fig:e_D_G}: 
Absolute relative error on (a) drag coefficient and (b) mean flow rate 
for $\delta=0.1$ (circles), $\delta=1$ (squares) and $\delta=10$ (triangles) 
versus the size $h/\delta$ of the physical grid. 
Lines are the least-mean square fit of the results. 
$N_v=8000$, $r_{v}=0.945$, $N_t=1024$ for $\delta=0.1$;
$N_v=8000$, $r_{v}=0.945$, $N_t=6144$ for $\delta=1$;
$N_v=5832$, $r_{v}=0.970$, $N_t=8192$ for $\delta=10$.

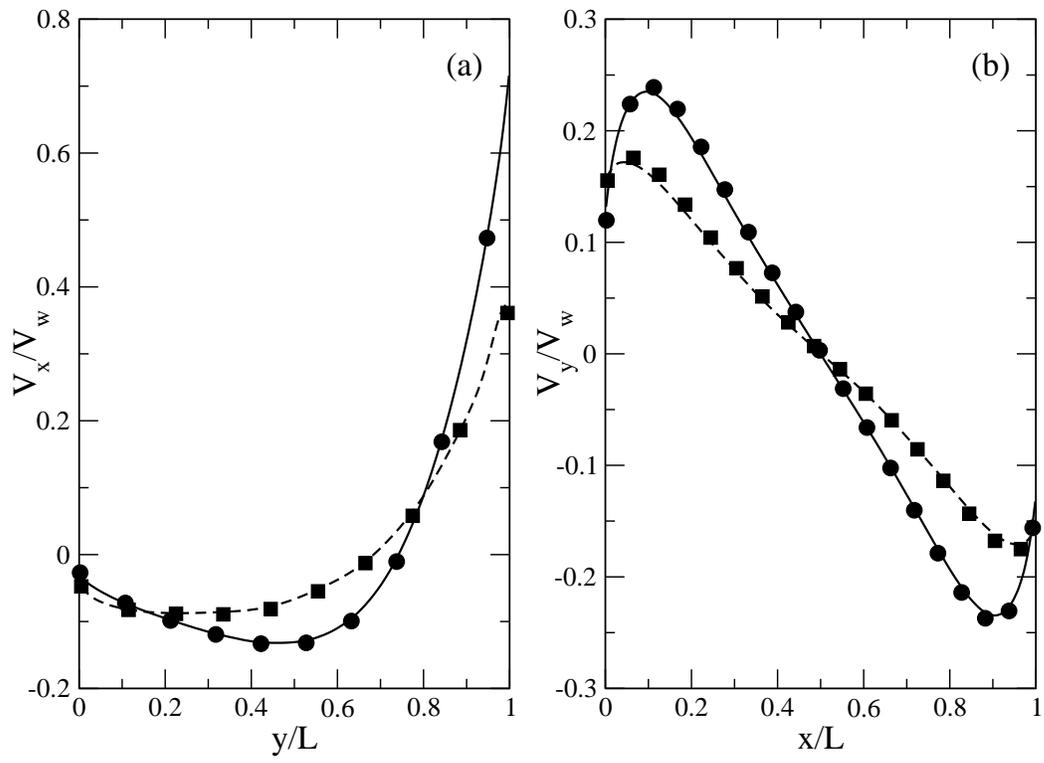
\begin{figure}[h!]
\begin{center}
\epsfig{file=Fig2.eps,height=10cm}
\caption{}
\label{fig:profili}
\end{center}
\end{figure}

\noindent
Fig. \ref{fig:profili}: 
Profiles of the dimensionless (a) horizontal mean velocity along the vertical
line crossing the center of the cavity, and (b) vertical mean velocity 
component along the horizontal line crossing the center of the main vortex.
Solid and dashed lines: numerical solutions obtained by solving 
Eq.~\eqref{eq:devBoltz} with the semi-regular method
for $\delta=10$ and $\delta=0.1$, respectively.
Circles and squares: numerical solutions reported in Ref.~\cite{VVS08} 
for $\delta=10$ and $\delta=0.1$, respectively. 
$N_r=25600$, $r_x=0.996$, $r_y=0.998$, $N_v=8000$, $r_{v}=0.945$, $N_t=1024$ 
for $\delta=0.1$;
$N_r=36864$, $r_x=0.990$, $r_y=0.995$, $N_v=5832$, $r_{v}=0.970$, $N_t=8192$ 
for $\delta=10$.

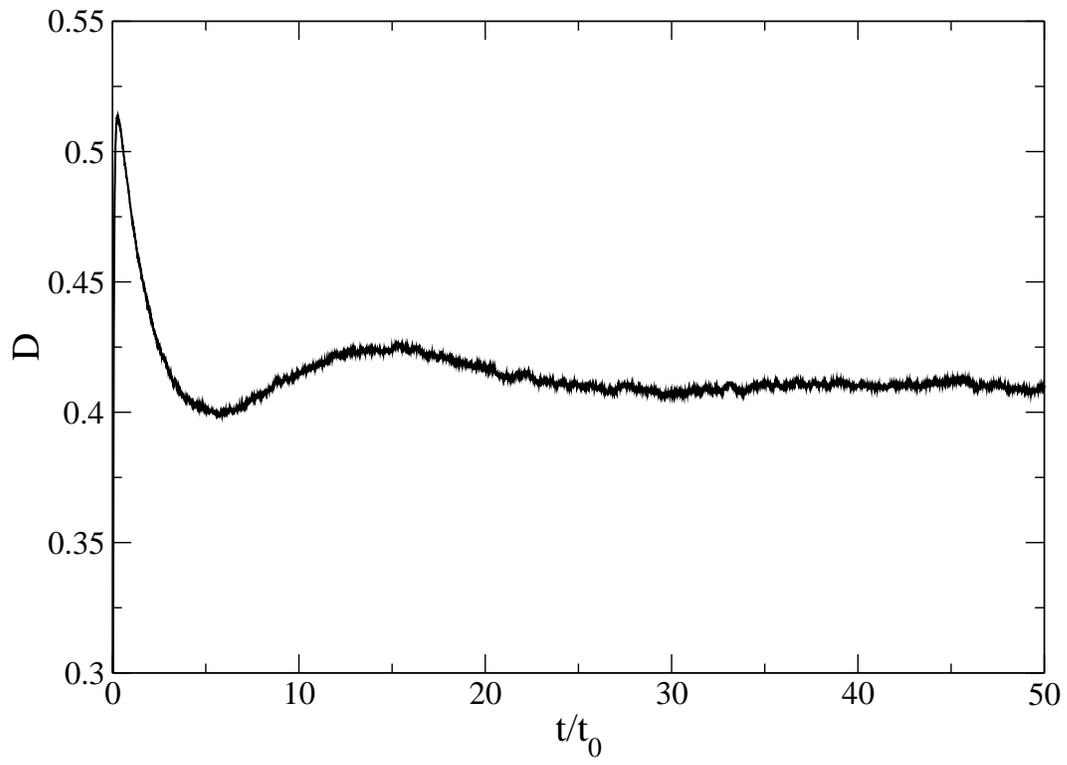
\begin{figure}[h!]
\begin{center}
\epsfig{file=Fig3.eps, height=10cm}
\caption{}
\label{fig:dragVsTime}
\end{center}
\end{figure}

\noindent
Fig. \ref{fig:dragVsTime}: 
Drag coefficient over the moving wall, $D$, versus 
dimensionless time, $t/t_{0}$. $\delta=1$.

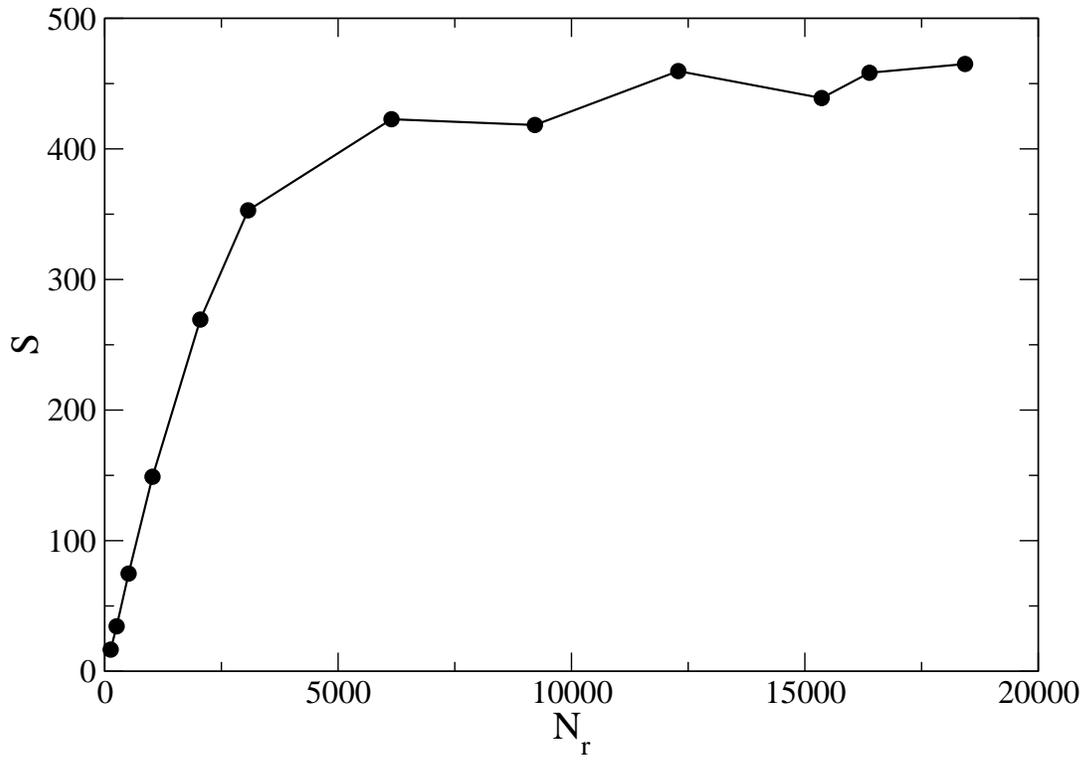
\begin{figure}[h!]
\begin{center}
\epsfig{file=Fig4.eps, height=10cm}
\caption{}
\label{fig:speedUp}
\end{center}
\end{figure}

\noindent
Fig. \ref{fig:speedUp}: 
Overall speed-up, $S$, versus the number of cells in the physical space,
$N_{r}$. 
$\delta=1$, $N_v=8000$, $N_t=6144$. 

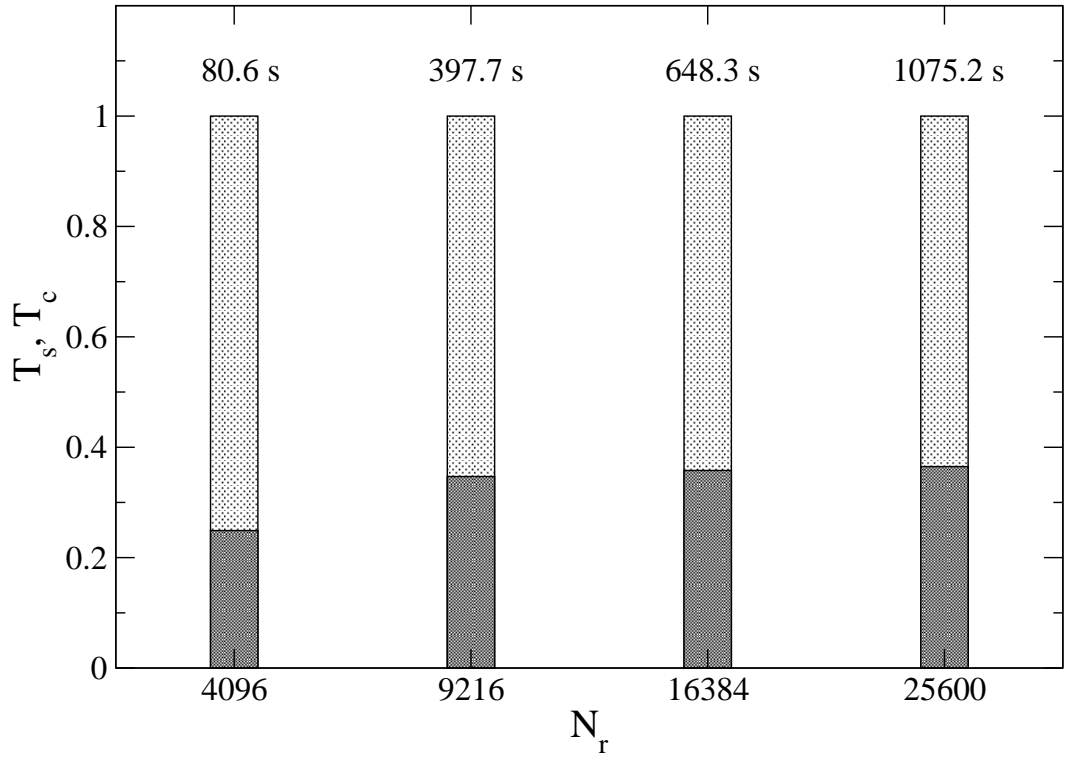
\begin{figure}[h!]
\begin{center}
\epsfig{file=Fig5.eps, height=10cm}
\caption{}
\label{fig:times}
\end{center}
\end{figure}

\noindent
Fig. \ref{fig:times}: 
Relative time spent on the streaming step (dark bar) and on
collision step (light bar). The numbers above the bars refer to the 
total execution time expressed in seconds.
$\delta=1$, $N_v=8000$, $N_t=6144$.

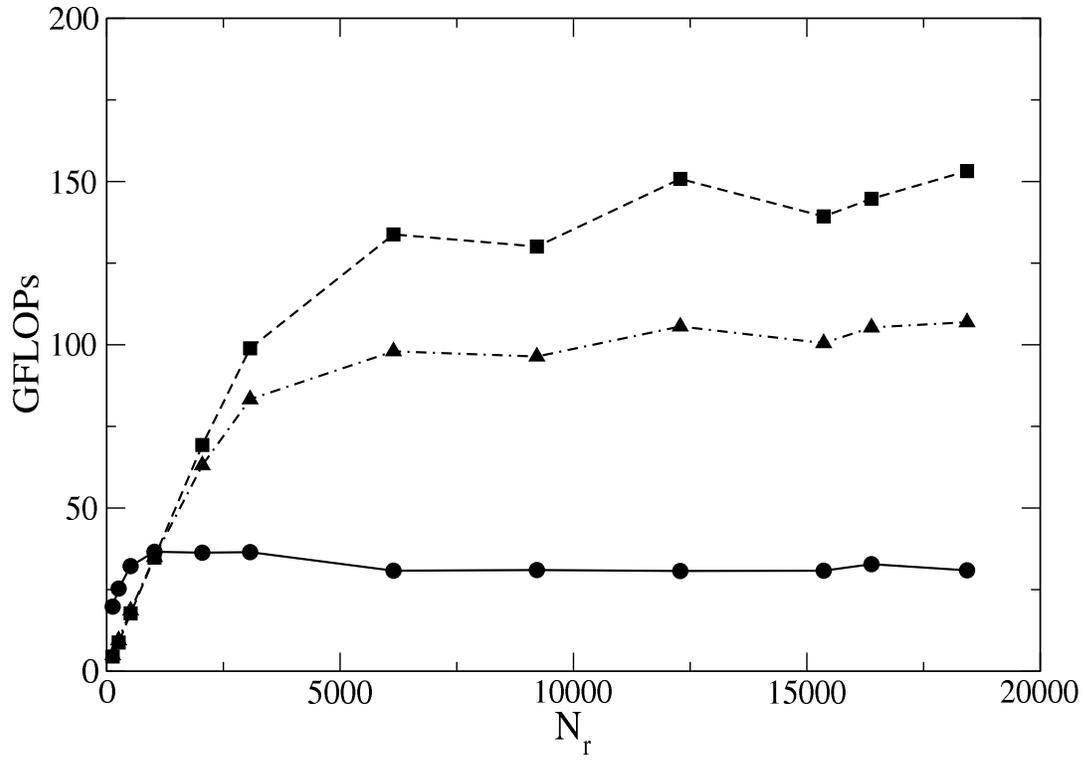
\begin{figure}[h!]
\begin{center}
\epsfig{file=Fig6.eps, height=10cm}
\caption{}
\label{fig:GflopDriCav}
\end{center}
\end{figure}

\noindent
Fig. \ref{fig:GflopDriCav}: 
GFLOPs versus the number of cells in the physical space, $N_{r}$. 
Solid line with circles: streaming step;
dashed line with squares: collision step;
dot and dashed line with triangles: overall code.
$\delta=1$, $N_v=8000$, $N_t=6144$.

\noindent
Table \ref{tab:DeG}:
Drag coefficient, $D$, and mean flow rate, $G$, versus the rarefaction
parameter, $\delta$. $D_e$ and $G_e$ represent the estimated exact values.

\newpage

\begin{algorithm}
\caption{GPU pseudo-code of the two-dimensional streaming
kernel} \label{alg:streaming2D}
\begin{algorithmic}[1]
 \REQUIRE $\mbox{Cu}_{x}$, Courant number along $x$ direction
 \REQUIRE$\mbox{Cu}_{y}$, Courant number along $y$ direction
 \REQUIRE $B_{x}$, number of threads in $x$ direction
 \REQUIRE $B_{y}$, number of threads in $y$ direction
 \REQUIRE $t_{x}$, thread index in $x$ direction within the block
 \REQUIRE $t_{y}$, thread index in $y$ direction within the block
 \REQUIRE $f_{\mathrm{sh}}$, matrix $(B_x+1)\times (B_y+1)$
           in the shared memory
 \FOR{$I_{by} = N_y/B_{y}-1$ to $0$}
 \FOR{$I_{bx}=N_x/B_{x}-1$to $0$}
 \STATE$i_{x} \leftarrow t_x + B_{x}I_{bx}$
 \STATE$i_{y} \leftarrow t_y + B_{y} I_{by}$
 \STATE$h_{\mathrm{sh}}(t_x+1,t_y+1) \Leftarrow h^{n}_{i_x,i_y;\bj}$
 \IF{$t_y == 0$}
 \IF{$i_y-1 < 0$}
 \STATE$h_{\mathrm{sh}}(t_x,t_y)\leftarrow \mathrm{boundaryFlux}$
 \ELSE
 \STATE $h_{\mathrm{sh}}(t_x,t_y)\Leftarrow h^{n}_{i_x,i_y-1,\bj}$
 \ENDIF
 \ENDIF
 \IF{$t_x == 0$} \IF{$i_x-1< 0$}
 \STATE$h_{\mathrm{sh}}(t_x,t_y)\leftarrow \mathrm{boundaryFlux}$
 \ELSE
 \STATE $h_{\mathrm{sh}}(t_x,t_y)\Leftarrow h^{n}_{i_x-1,i_y,\bj}$
 \ENDIF
 \ENDIF
 \STATE syncthreads
 \STATE$h_{\mathrm{rg}}\leftarrow
 (1-\mbox{Cu}_x-\mbox{Cu}_y) h_{\mathrm{sh}}(t_x+1,t_y+1) +
        \mbox{Cu}_y\,h_{\mathrm{sh}}(t_x,t_y+1) + 
        \mbox{Cu}_x\,h_{\mathrm{sh}}(t_x+1,t_y)$
 \STATE $h_{\mathrm{rg}} \Rightarrow
\tilde{h}^{n+1}_{i_x,i_y,\bj}$
 \STATE$I_{bx} \leftarrow I_{bx}-1$
 \ENDFOR
 \STATE$I_{by} \leftarrow I_{by} -1$
 \ENDFOR
\end{algorithmic}
\end{algorithm}

\newpage

\begin{algorithm}
\caption{GPU pseudo-code of the sampling kernel}
\label{alg:sampling}
\begin{algorithmic}[1]
 \REQUIRE $i$, global thread index in the grid
 \STATE $\bv\leftarrow$BoxMulller
 \STATE $\bv_{1} \leftarrow$BoxMulller
 \STATE $\bkhat \leftarrow$UnitSphere
 \STATE$\bv_{r} \leftarrow \bv_{1}-\bv$
 \STATE$\bv^{*} \leftarrow \bv + (\bv_{r} \cdot \bkhat)\bkhat$
 \STATE$\bv^{*}_{1} \leftarrow \bv_{1} - (\bv_{r} \cdot \bkhat)\bkhat$
 \STATE cells$(\bv)\Rightarrow I(i)$
 \STATE cells$(\bv_{1})\Rightarrow I_{1}(i)$
 \STATE cells$(\bv^{*})\Rightarrow I^{*}(i)$
 \STATE cells$(\bv^{*}_{1})\Rightarrow I^{*}_{1}(i)$
 \STATE $g_{\bj} \leftarrow \pi \sigma^2 n_{0}^{2}
 (\Delta \mathcal{V}_{\bj} \Phi_{\bj})^{-1} 
|\bv_{r}\cdot \bkhat| \Delta t$
 \STATE $g_{\bj_{1}} \leftarrow \pi \sigma^2 n_{0}^{2}
 (\Delta \mathcal{V}_{\bj_{1}} \Phi_{\bj_{1}})^{-1} 
|\bv_{r}\cdot \bkhat| \Delta t$
 \STATE$g_{\bj^{*}} \leftarrow \pi \sigma^2 n_{0}^{2}
 (\Delta \mathcal{V}_{\bj^{*}} \Phi_{\bj^{*}})^{-1}
|\bv_{r}\cdot \bkhat| \Delta t$
 \STATE $g_{\bj^{*}_{1}} \leftarrow \pi \sigma^2 n_{0}^{2}
 (\Delta \mathcal{V}_{\bj^{*}_{1}} \Phi_{\bj^{*}_{1}})^{-1}
|\bv_{r}\cdot \bkhat| \Delta t$
 \STATE $g_{\bj} \Rightarrow C(i)$
 \STATE $g_{\bj_{1}} \Rightarrow C_{1}(i)$
 \STATE$g_{\bj^{*}} \Rightarrow C^{*}(i)$
 \STATE $g_{\bj^{*}_{1}} \Rightarrow C^{*}_{1}(i)$
\end{algorithmic}
\end{algorithm}

\begin{algorithm}
\caption{GPU pseudo-code of the collision kernel}
\label{alg:collision}
\begin{algorithmic}[1]
 \REQUIRE $\bi$, global thread index in the grid
 \FORALL{${\bj}$}
 \STATE $h \Leftarrow \hat{h}^{n+1}_{\bi,\bj}$
 \STATE $\rho \leftarrow \rho + \Phi_{0,\bj} \, h $
 \STATE $\bu \leftarrow \bu + \bv_{\bj} \; \Phi_{0,\bj} \, h $
 \STATE $e \leftarrow e + |\bv_{\bj}|^{2}\;\Phi_{0,\bj} \, h $
 \ENDFOR
 \FOR{$m = 1$ to $N_t$}
 \STATE $h \Leftarrow \tilde{h}^{n+1}_{\bi,I(m)}$
 \STATE $h_{1} \Leftarrow \tilde{h}^{n+1}_{\bi,I_1(m)}$
 \STATE $h^{*} \Leftarrow \tilde{h}^{n+1}_{\bi,I^{*}(m)}$
 \STATE $h^{*}_{1} \Leftarrow \tilde{h}^{n+1}_{\bi,I^{*}_{1}(m)}$
 \STATE $g \leftarrow h + h_1 + \epsilon \, h \,h_1$
 \STATE $h \leftarrow h - C(m) \; g$
 \STATE $h_{1} \leftarrow h_{1} - C_{1}(m)\; g$
 \STATE $h^{*} \leftarrow h^{*} + C^{*}(m)\; g$
 \STATE $h^{*}_{1} \leftarrow h^{*}_{1} + C_{1}^{*}(m)\; g$
 \STATE $h \Rightarrow \tilde{h}^{n+1}_{\bi,I(m)}$
 \STATE $h_{1} \Rightarrow \tilde{h}^{n+1}_{\bi,I_1(m)}$
 \STATE $h^{*} \Rightarrow \tilde{h}^{n+1}_{\bi,I^{*}(m)}$
 \STATE $h^{*}_{1} \Rightarrow \tilde{h}^{n+1}_{\bi,I^{*}_{1}(m)}$
 \ENDFOR
 \FORALL{${\bj}$}
 \STATE $h \Leftarrow \tilde{h}^{n+1}_{\bi,\bj}$
 \STATE $a_{11} \leftarrow a_{11} + \Phi_{0,\bj} \, h $
 \STATE $a_{12} \leftarrow a_{12} + {\bv}_{\bj} \; \Phi_{0,\bj} \, h$
 \STATE $a_{13} \leftarrow a_{13} + |{\bv}_{\bj}|^{2}\;\Phi_{0,\bj} \,h$
 \STATE $\vdots$
 \STATE // others moments of the distribution function
 \STATE$\vdots$
 \ENDFOR
 \STATE [$A,\bB,C$]=linearSolver($n,\bu,e,a_{11},a_{12},a_{13},\ldots$)
 \FORALL{${\bj}$}
 \STATE $h \Leftarrow \tilde{h}^{n+1}_{\bi,\bj}$
 \STATE $h \leftarrow 1/\epsilon(1+\epsilon h)(1+A+\bB \cdot\bv_{\bj} + 
         C \bv^{2}_{\bj}-1)$
 \STATE $h \Rightarrow h^{n+1}_{\bi,\bj}$
 \STATE $\vdots$
 \STATE // compute macroscopic quantities
 \STATE $\vdots$
\ENDFOR
\end{algorithmic}
\end{algorithm}

\newpage

\newpage

\newpage

\newpage

\newpage

\newpage

\newpage

\begin{table}[h!]
\begin{center}
\begin{tabular}{|c||c|c|c|c|c|c||c|c|c|c|}
\hline
$\delta$ & $D$ & $D_e$ & $D$ (Ref.~\cite{VVS08}) & 
           $G$ & $G_e$ & $G$ (Ref.~\cite{VVS08}) \\ \hline \hline
 0.1     & 0.6712 & 0.6815 & 0.676-0.678  
         & 0.0955 & 0.0977 & 0.0973-0.0976 \\ \hline
 1       & 0.6266 & 0.6389 & 0.625-0.631 
         & 0.1017 & 0.1039 & 0.104-0.105  \\ \hline
 10      & 0.4096 & 0.4176 & 0.412-0.415 
         & 0.1427 & 0.1451 & 0.145-0.145 \\ \hline 
\end{tabular}
\caption{}
\label{tab:DeG}
\end{center}
\end{table}


\begin{thebibliography}{50}
\bibitem{c88} C. Cercignani,
              The Boltzmann Equation and Its Applications,
                Springer-Verlag, New York, 1988.
\bibitem{b94} G. A. Bird,
              Molecular Gas Dynamics and the Direct Simulation of Gas
              Flows,
              Oxford University Press, 1994.
\bibitem{fgl05} A. Frezzotti, L. Gibelli, S. Lorenzani,
                ``Mean field kinetic theory description of evaporation of a
                  fluid into vacuum'',
                 Phys. Fluids 17, 012102-12 (2005).
\bibitem{hh07} T. M. M. Homolle, N. G. Hadjiconstantinou,
                ``A low-variance deviational simulation Monte Carlo for the
                  Boltzmann equation'',
               J. Comput. Phys. 226 (2007) 2341-2358.
\bibitem{w08} W. Wagner,
               ``Deviational particle Monte Carlo for the Boltzmann equation'',
                Monte Carlo Methods and Applications 14 (2008)
                191-268.
\bibitem{f91} A. Frezzotti,
               ``Numerical study of the strong evaporation of a binary
                 mixture'',
               Fluid Dynamics Research 8 (1991) 175-187.
\bibitem{t05} F. Tcheremissine,
                ``Direct numerical solution of the Boltzmann Equation'',
                AIP Conf. Proc. 762 (2005)
                677-685.
\bibitem{a01} V. V. Aristov,
              Direct Methods for Solving the Boltzmann Equation and Study
              of Nonequilibrium Flows,
              Springer-Verlag, New York, 2001.
\bibitem{f07} A. Frezzotti,
              ``A numerical investigation of the steady evaporation of a
                polyatomic gas'',
              Eur. J. Mech. B: Fluids 26 (1) (2007) 93-104.
\bibitem{mbkj09} P. Matinsen, J. Blaschke, R. K\"unnemeyer, R. Jordan,
               ``Accelerating Monte Carlo simulations with an \nvidia
                 \, graphics processor'',
               Comput. Phys. Commun. 180 (10) (2009) 1983-1989. 
\bibitem{jk10} M. Januszewski, M. Kostur,
               ``Accelerating numerical solution of stochastic differential
                 equations with cuda'',
               Comput. Phys. Commun. 181 (1) (2010) 183-188. 
\bibitem{fgg10} A. Frezzotti, G. P. Ghiroldi, L. Gibelli,
                ``Solving Model Kinetic Equations on GPUs'',
                arxiv:0903.4044v1
\bibitem{bh08} L. L. Baker, N. G. Hadjiconstantinou,
               ``Variance-reduced Monte Carlo solutions of the Boltzmann 
                 equation for low-speed gas flows: A discontinuous Galerkin 
                 formulation'', Int. J. Numer. Meth. Fluids 58 (2008)
                 381-402
\bibitem{at80} V.V. Aristov and F.G. Tcheremissine,
               U.S.S.R. Comput. Math. Phys. 20 (1980) 208-225.
\bibitem{n08} NVIDIA Corporation,
                 ``NVIDIA CUDA Programming Guide'',
                 Jun. 2008. Version 2.0. http://www.nvidia.com/CUDA
\bibitem{M09} P. Micikevivius,
              ``3D finite difference computation on GPUs using CUDA'',
              ACM International Conference Proceeding Series, 383 (2009)
              79-84.
\bibitem{VVS08} S. Varoutis, D. Valougeorgis, F. Sharipov,
               ``Application of the integro-moment method to steady-state
                 two-dimensional rarefied gas flows subject to boundary
                 induced discontinuities'',
                J. Comput. Phys. 227 (2008) 6272-6287.
\bibitem{cc90} S. Chapman, T. G. Cowling,
               The mathematical theory of non-uniform gases,
               Cambridge University Press, 1990.
\bibitem{s06} M. D. Salas,
              ``Some observations on grid convergence'',
               Comp. \& Fluids 35 (2006) 688-692.
\bibitem{eld08} E. Elsen, P. LeGresley, E. Darve,
               ``Large calculation of the flow over a hypersonic vehicle
                 using a GPU'',
                J. Comp. Phys. 227 (2008)
                10148-10161.
\bibitem{alt08} J. A. Anderson, C. D. Lorenz, A. Travesset,
                ``General purpose molecular dynamics simulations fully
                  implemented on graphics processing units'',
                J. Comp. Phys. 227 (2008)
                5342-5359.
\end{thebibliography}
\end{document}